\documentstyle[12pt,epsfig]{article}
\title{
Primary Cosmic-ray Spectrum and\\
the Intensity of Atmospheric Neutrinos\thanks{Talk given at the
Satellite Conference, "New Era in Neutrino Physics",
Tokyo, 11-12 June, 1998.  Work supported in part by the U.S. Department
of Energy.}}

\author{Thomas K. Gaisser\\
 Bartol Research Institute, University of Delaware\\
Newark, DE 19716 USA}
\begin{document}
\maketitle

\section*{Abstract}

Measurements of the intensity of primary cosmic-ray protons reported
in the last two years suggest a lower normalization than some
earlier measurements.  Here we comment on the measurements, compare
them to the assumptions made in two independent calculations of
atmospheric neutrinos and discuss the possible consequences for
interpretation of measurements of atmospheric neutrinos.

\vspace{1cm}

In a series of recent papers~\cite{SK1,SK2,SK3}, the
Super-Kamiokande collaboration has presented evidence for
neutrino oscillations based on a high-statistics
sample of neutrino interactions
in their large detector.  The importance of
these results, which consolidate and extend the previous discovery
of an ``atmospheric neutrino anomaly''~\cite{IMB,Kam,Soudan,MACRO},
calls for a new look at
the calculation of the flux of atmospheric neutrinos, which is
the starting point for interpretation of the data.
I give a general analysis of the neutrino flux calculation
elsewhere~\cite{nu98}.  Here I consider only the primary spectrum,
which affects mainly the normalization of the neutrino flux.

Neutrino interactions in Super-K are divided into sub-GeV~\cite{SK1}
and multi-GeV~\cite{SK2} events.  The most numerous are the
sub-GeV events ($E_\nu\sim 1$~GeV), which come from primary
cosmic rays with energy between 5 and 100 GeV/nucleon.
The multi-GeV sample corresponds very roughly to a factor 5 to 10
higher neutrino energy and hence to a correspondingly higher
range of primary energy.  Neutrino-induced muons that enter
the detector from below and stop cover a similar energy range
to the multi-GeV sample.  Finally, upward, through-going muons
as measured by MACRO~\cite{MACRO} and several other underground
detectors, including the water detectors and Baksan~\cite{Baksan},
correspond to neutrino energies from $\sim 10$ to $\sim 10^4$~GeV and hence
to primary energies up to $\sim10^5$~GeV.
For any one sample, uncertainty in the primary spectrum affects
mainly the normalization.  Because of the long lever arm in energy,
however, a small uncertainty in the slope can have a significant
effect on the calculated ratio of, for example, stopping to through-going
muons.

The standard reference for the primary spectrum in the energy
region responsible for the sub-GeV neutrino event sample
was the work of Webber {\it et al.} \cite{Webber} until 1991.
This was based on two balloon flights, one in 1976 and one in 1979,
in which the spectra of hydrogen and helium were measured and compared.
The LEAP experiment \cite{LEAP} flew in 1987 during solar minimum.
It covered a similar energy range with the same spectrometer in a
somewhat different configuration.  The LEAP experiment
gave results with a normalization
for protons about 50\% lower than the Webber results.
This difference is larger than the uncertainty of either measurement
and therefore implies the existence of systematic uncertainties that
have not been fully understood.  Since the atmospheric neutrino
flux is proportional to the normalization of the primary spectrum,
there is a corresponding ambiguity in the calculated flux of
atmospheric neutrinos.

This was the situation\footnote{There was also a
flight of the MASS instrument but it occurred on September 5, 1989
during a record-breaking Forbush decrease, so it corresponds
to an extremely high degree of solar modulation, and its
interpretation is correspondingly complicated.  In particular,
its normalization is tied to that of the earlier measurement
of Ref. \cite{Webber}.} 
from 1991 until about two years ago,
when a series of magnetic spectrometer
experiments started to appear in print.  (See Table \ref{tab1}.)  The
results of these more recent experiments support the lower normalization
of the LEAP measurement.    The primary goal of these experiments
was to measure the spectrum of cosmic-ray antiprotons.  In the process,
they necessarily measured the primary spectrum of protons and in
most cases also of helium.  Some of them also measured the spectrum
of muons in the atmosphere during ascent as well as on the ground
and at float altitude.   The measured muon spectra are very closely
related to atmospheric neutrino fluxes \cite{Perkins}.

The IMAX \cite{IMAX} and
CAPRICE \cite{CAPRICE} groups used modified versions of the same magnetic
spectrometer as Webber {\it et al.} and LEAP, but they were designed to
provide two independent measurements of the trajectory through
the spectrometer so that
efficiencies could be determined with greater certainty.
The most recent result is the
BESS measurement \cite{BESS} reported at this meeting.
This experiment uses a completely different spectrometer
and magnet with cylindrical geometry \cite{BESS2}.  The BESS results also
are in good agreement with the LEAP measurements.

\begin{table}[t]
   \caption{Measurements of primary cosmic-rays and atmospheric muons}
\vspace{.5pc}
\begin{center}
\begin{tabular}{l|ccc} \hline
 & year flown & Primary & Muons? \\ \hline
Webber & '76,'79 & p,He \cite{Webber} & no \\
LEAP & '87 & p,He \cite{LEAP} & no \\
MASS & '89,'91 & p,He \cite{MASSp} & yes\cite{MASS1,MASS2} \\
IMAX & '92 & p,He \cite{IMAX} & yes \cite{IMAXmu} \\
CAPRICE & '94 & p,He \cite{CAPRICE} & yes \cite{CAPmu}\\
BESS & '97 & p \cite{BESS} & not published \cite{BESSmu} \\
HEAT & '94,'95 & & yes \cite{HEAT}\\ \hline
\end{tabular}
\label{tab1}
\end{center}
\end{table}

A graphical summary of the BESS spectrum of hydrogen, together with
a comprehensive compilation of the previous measurements is given in
Ref.~\cite{BESS}.  
Fig. 1 here shows another compilation of the measurements of primary spectra
of nuclei in which the assumptions of two calculations of the neutrino flux
\cite{AGLS,HKKM} are also shown.  The (rather slight)
consequences of the difference in
slope between \cite{AGLS} and \cite{HKKM} up to $\sim100$~GeV
are discussed in~\cite{nu98}.  The difference in normalization
is a bigger effect.  In the absence of all other differences
in input, the rate of sub-GeV interactions of $\nu_\mu+\bar{\nu}_\mu$
would be 12\% higher using the flux of Ref.~\cite{HKKM} than
with the flux used by \cite{AGLS}.  In fact, the expected rates
differ by less than 5\%~\cite{SK1}.  This accidental cancellation
is a consequence of the fact that the yields of pions, and hence
of muons and neutrinos, in interactions of protons with nuclei
of the atmosphere are higher in Ref. \cite{AGLS} than in \cite{HKKM}.
This in turn may to some extent reflect the fact that both
calculations are also trying to fit the same measurements of muons
\cite{MASS1} high in the atmosphere.

\vspace{0.8cm}

\noindent
{\bf \large Conclusions.}

\begin{itemize}
\item The ratio of observed to calculated sub-GeV electron-like
events is  1.21, and the corresponding ratio for
muon-like events is 0.74 with a calculation based on the calculation
of the neutrino flux of Honda {\it et al.}~\cite{HKKM} (analysis A
of Ref.~\cite{SK1}).  If the
primary flux in that calculation were reduced to fit the recent
data above 10 GeV, with no other changes being made, the excess of
electron neutrinos would be increased by a similar amount
(and the deficit of muon neutrinos correspondingly reduced).
\item The corresponding numbers based on the neutrino flux calculation
of Ref.~\cite{AGLS} (analysis B of Ref.~{SK1}) are essentially the
same, 1.18 and 0.76.  Here, however, renormalizing Ref.~\cite{AGLS}
to the lower primary
spectrum measurements would have a smaller ($<$5\%) effect because
the assumed primary spectrum of protons is already fairly low.
\item Note that Fig. 1 shown here differs from the figure that I showed
at the conference, which was Fig. 2 from Ref.~\cite{AGLS}.
That figure is wrong in the sense that the fit it shows for hydrogen
is not the hydrogen spectrum that was actually used in the calculation
of the neutrino flux.  An erratum is in preparation~\cite{erratum}.
The correct spectrum is shown as the dashed line in Fig. 1 
in this paper.
\item Webber \cite{private} points out that the low normalization
implied by the results of Refs.~\cite{LEAP,IMAX,CAPRICE,BESS} may be 
inconsistent with simple measurements of the integral spectrum
made without spectrometers by
using the local geomagnetic field to set the minimum energy.
\item Several components enter into the normalization of the calculated
atmospheric neutrino flux in the region of energy most important
for sub-GeV events, most notably the yield of pions in collisions
of protons with nitrogen and oxygen in the atmosphere and the normalization
of the primary spectrum.  The comparison between Ref.~\cite{AGLS}
and Ref.~\cite{HKKM} illustrates that the two uncertainties may
cancel and thus obscure the systematic uncertainties in the calculations.
\item Measurements of muons high in the
atmosphere provide an independent constraint that is in principle
somewhat more direct \cite{Perkins}.
The HEAT measurement of the atmospheric muon flux is
somewhat lower that the MASS measurement, which has been compared
to the corresponding neutrino flux calculations~\cite{AGLS,HKKM}.
\item An independent calculation, now in preliminary stages~\cite{Battistoni},
has pion yields that are somewhat lower than those used in Ref.~\cite{AGLS}.
\end{itemize}

In view of the uncertainties itemized above (and others discussed in
Ref.~\cite{nu98}), the anomalous flavor ratios and 
the anomalous angular dependence  observed by Super-K~\cite{SK3}
remain the cleanest evidence for neutrino oscillations because
the normalization uncertainty cancels.
On the other hand, it seems unlikely that improvements or corrections
to the calculation will lead to much increase in the predicted
neutrino flux unless it turns out that the higher normalization
of the primary spectrum is after all correct.

\noindent
{\bf Acknowledgements.} I am grateful for helpful conversations
with Bill Webber, Todor Stanev, Shuji Orito, T. Sanuki, Per Carlson,
Tom Francke and Mirko Boezio.

\newpage

\begin{figure}[!htb]
\centerline{\epsfig{figure=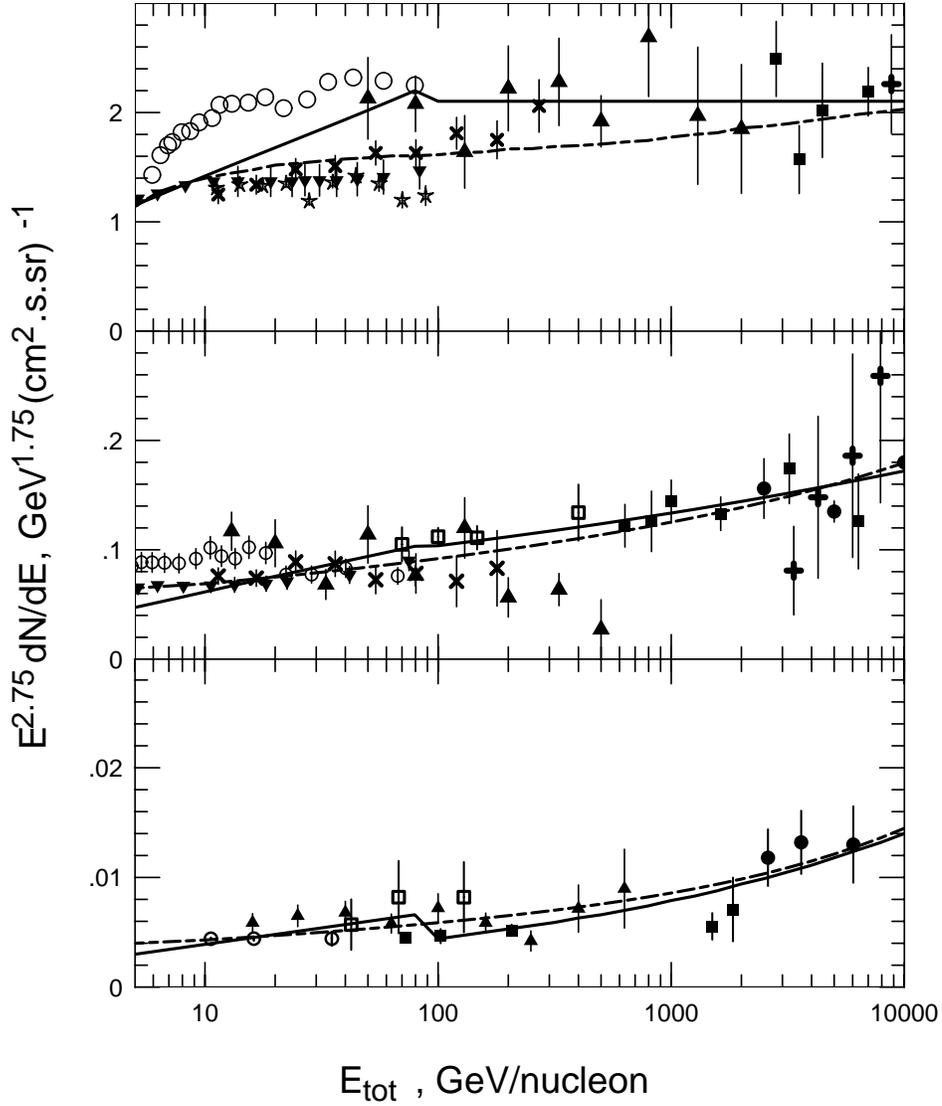,height=15cm}}
\caption{Spectra of hydrogen (top panel), helium (middle panel) and
CNO nuclei (bottom panel).  Open circles show the Webber {\it et al.}
data as quoted in the IMAX paper~\protect\cite{IMAX}.  Inverted, filled
triangles show LEAP; X show IMAX~\protect\cite{IMAX}, and
stars show CAPRICE~\protect\cite{CAPRICE}.  Higher energy data are obtained
with calorimeters.  Other references are given in Ref. \protect\cite{AGLS}.
Dashed curves show the primary spectra of Ref.~\protect\cite{AGLS} and solid
curves those of Ref.~\protect\cite{HKKM}.
}
\end{figure}

\newpage
%%%%%%%%%%%%%%%%%%%%%%%%%%%%%%%%%%%%%%
%       Please fill out items listed below. See %%% EXAMPLE %%%
%
%%%%%%%%%%%%%%%%%%%%%%%%%%%%%%%%%%%%%%


\newpage

\begin{thebibliography}{9}
\bibitem {SK1} Y. Fukuda {\it et al.}, Phys. Lett. B433 (1998) 9.
\bibitem {SK2} Y. Fukuda {\it et al.}, hep-ex/9805006 (to be published).
\bibitem {SK3}  Y. Fukuda {\it et al.}, Phys. Rev. Letters 81 (1998) 1562.
\bibitem {IMB} R. Becker-Szendy {\it et al.} Phys. Rev. D46 (1992) 3720
and references therein.
\bibitem {Kam}  Y. Fukuda {\it al.} (Kamiokande Collaboration)
Phys. Lett. B335 (1994) 237 and references therein.
\bibitem{Soudan} W.W.M. Allison {\it et al.} (Soudan Collaboration),
Phys. Lett. B391 (1997) 491.
\bibitem{MACRO} M. Ambrosio {\it et al.} (MACRO Collaboration),
hep-ex/9807005.  For stopping events in MACRO see M. Spurio,
hep-ex/9808001.
\bibitem {nu98} T.K. Gaisser in Proc. Neutrino98 Conference,
Takayama, to appear in Nucl. Phys. B (Conf. Proc.).
\bibitem {Baksan} S. Mikheyev, 5th TAUP Workshop proceedings,
Gran Sasso, Italy (1997). 
\bibitem {Webber} W.R. Webber, R.L. Golden \& S.A. Stephens,
Proc. 20th Int. Cosmic Ray Conf. (Moscow) vol. 1 (1987) 325.
\bibitem {LEAP} E.S. Seo {\it et al.}, Ap. J. 378 (1991) 763.
\bibitem{Perkins}  D.H. Perkins, Astroparticle Physics 2 (1994) 249.
\bibitem {IMAX} W. Menn, {\it et al.}, Proc. 25th Int. Cosmic Ray
Conf. (Durban) vol. 3 (1997) 409.
\bibitem {CAPRICE} G. Barbiellini, {\it et al.}, Proc. 25th Int. Cosmic Ray
Conf. (Durban) vol. 3 (1997) 369.  Also, M. Boezio {\it et al.}, (1998) 
(to be published).
\bibitem {BESS} S. Orito, {\it et al.}, talk presented by T. Sanuki
at this meeting.
\bibitem {BESS2} H. Matsunaga {\it et al.},
Proc. 25th Int. Cosmic Ray Conf. (Durban) vol. 4 (1997) 205.
\bibitem {MASSp} W.R. Webber {\it et al.}, Ap. J. 380 (1991) 230.
\bibitem{MASS1} R. Bellotti, {\it et al.}, Phys. Rev. D53 (1996) 35.
See also M.P. De Pascale {\it et al.}, J. Geophys. Res. 98 (1993) 3501
for a measurement from the ground.
\bibitem{MASS2} G. Basini {\it et al.}, Proc. 25th Int. Cosmic Ray Conf.
(Durban) vol. 6 (1997) 381.
\bibitem{IMAXmu} J.F. Krizmanic {\it et al.} Proc. 24th Int. 
Cosmic Ray Conf. (Rome) vol. 1 (1995) 593.
\bibitem{CAPmu} G. Barbiellini {\it et al.}, Proc. 25th Int. Cosmic Ray Conf.
(Durban) vol. 6 (1997) 317.  (This reference is the charge ratio at the
top of the atmosphere.  Results of measurements of muons during
ascent are forthcoming--P. Carlson {\it et al.}, private communication).
\bibitem{BESSmu} Measurements of the muon spectrum have been made
with the BESS detector on the ground at Lynn Lake and at Tsukuba
(S. Orito, private communication).
\bibitem{HEAT} G. Tarl\'{e} {\it et al.}, Proc. 25th Int. Cosmic Ray Conf.
(Durban) vol. 6 (1997) 321.  S. Coutu {\it et al.}, Proc. Int. Conf. on
High Energy Physics (Vancouver, 1998, to be published).
\bibitem{AGLS} Vivek Agrawal, T.K. Gaisser, Paolo Lipari and Todor Stanev,
Phys. Rev. D53 (1996) 1314.
\bibitem{HKKM} M. Honda, T. Kajita, K. Kasahara \& S. Midorikawa,
Phys. Rev. D52 (1995) 4985.
\bibitem{erratum} T.K. Gaisser \& Todor Stanev, erratum to be submitted
to PRD.
\bibitem{private} W. Webber, private communication.
\bibitem {Battistoni} G. Battistoni {\it et al.},
5th TAUP Workshop proceedings, Gran Sasso, Italy (1997).  
\end{thebibliography}
\end{document}